\documentclass[%
 reprint,
]{revtex4-2}

\usepackage{graphicx}
\usepackage{dcolumn}
\usepackage{bm}

\usepackage{amsthm,amsmath,amsfonts,amssymb}
\usepackage{mathtools,mathrsfs}
\usepackage{setspace}
\usepackage{xcolor}
\usepackage{thmtools} 

\usepackage[pdfdisplaydoctitle,colorlinks,breaklinks,urlcolor=blue,linkcolor=blue,citecolor=blue]{hyperref} 
\usepackage[nameinlink]{cleveref}

\newcommand{\R}{\mathbb{R}}

\newcommand{\T}{\mathbb{T}}

\newcommand{\Z}{\mathbb{Z}}

\newcommand{\set}[1]{\left\{#1\right\}}
\newcommand{\pa}[1]{\left(#1\right)}

\newcommand{\brak}[1]{\left\langle#1\right\rangle}

\theoremstyle{remark}

\numberwithin{equation}{section}

\begin{document}

\preprint{APS/123-QED}

\title[Point Vortices Decay of Correlations]{Decay of Time Correlations in Point Vortex Systems}

\author{Francesco Grotto}
 \email{francesco.grotto at unipi.it}
\affiliation{
 Dipartimento di Matematica, Universit\`a di Pisa\\ Largo Pontecorvo 5, 56127 Pisa, Italia
}

\author{Silvia Morlacchi}
\email{silvia.morlacchi at sns.it}
\affiliation{
 Scuola Normale Superiore, Classe di Scienze\\ Piazza dei Cavalieri, 7, 56126 Pisa, Italia
}

\date{\today}

\begin{abstract}
The dynamics of a large point vortex system whose initial configuration consists in uniformly distributed independent positions is investigated.
Time correlations of local observables of the vortex configuration are shown to be compatible with power law decay $1/t$,
providing additional insight on ergodicity and mixing properties of equilibrium dynamics in point vortex models.
\end{abstract}


\maketitle

\section{Introduction}

The main open problem in the context of incompressible 2D Euler dynamics is the long time behavior of the fluid.
The formation of coherent structures and self-organization of the fluid at large scales is a crucial feature of 2D turbulence \cite{Tabeling2002}, and it is intimately linked to quantitatively observable phenomena such as the inverse cascade in the energy spectrum \cite{Boffetta2012,Eyink2006}, anomalous dissipation of energy \cite{Eyink1994,Eyink2001} and irreversible mixing \cite{Dolce2022}.

In the present paper we analyze the temporal structure of equilibrium dynamics for a classical model in 2D fluid mechanics, 
that is the point vortex (PV) system, and we exhibit evidence of persistence in time correlations, 
in the form of power law decay of the latter.

Introduced by Helmholtz in 1857 \cite{Helmholtz1858}, PVs are widely known as the fluid discretization method adopted by Onsager \cite{Onsager1949} in laying grounds for the statistical mechanical approach to 2D turbulence.
PV methods have also found relevant applications in the numerical approximation of 2D incompressible fluids: important examples include boundary effects \cite{Chorin1973,cottet2000,mimeau2021,grotto2023uniform} and Kelvin-Helmholtz or Rayleigh–Taylor instabilities \cite{Krasny1986,ZHOU20171I,ZHOU20171II,Thalabard2020}.

The dynamics consists of a system of $N$ first-order singular ODEs, describing the evolution of points at which vorticity, 
\emph{i.e.} the curl of velocity, is ideally concentrated. For the sake of dealing with a finite reference measure, 
and in order to neglect boundary effects, we consider such system in a periodic domain $\T=[-1,1]^2$, 
\begin{equation}\label{eq:pv}\tag{PV}
    \dot x_i=\sum_{j\neq i}^N \gamma_j K(x_i-x_j),\quad i=1,\dots,N,
\end{equation}
in which the vector field induced by the interaction of couple of PVs,
$K(x_i-x_j)$, is given by 
\begin{gather*}
    K(x)=(-\partial_{x_2},\partial_{x_1})G(x), \\
    -(\partial_{x_1}^2+\partial_{x_2}^2) G(x)=\delta_0(x)-1,
\end{gather*}
where $G$ is the zero-averaged fundamental solution of Laplace's equation on $\T$.

The vorticity distribution
\begin{equation*}
    \omega(x)=\sum_{i=1}^N \gamma_i (\delta_0(x-x_i(t))-1)
\end{equation*}
formally satisfies the 2D Euler equations in vorticity form,
\begin{equation}\label{eq:2deuler}
    \begin{cases}
        \partial_t \omega + (u\cdot \nabla)\omega=0,\\
        \nabla\cdot u=0,
    \end{cases}
\end{equation}
under periodic boundary conditions on $\T=[-1,1]^2$, where $u(x)=\int K(x-y)\omega(y)dy$ by the Biot-Savart law.
Rescaling vortex intensities $\gamma_1,\dots,\gamma_N\in\R\setminus\set{0}$ 
(corresponding to velocity circulation around a single vortex) allows to fix a time scale for the dynamics: 
in our discussion we will consider a system of identical vortices and set $|\gamma_i|$ to unit.

The system in \cref{eq:pv} is not well-posed, as there exist initial configurations of arbitrarily 
many PVs leading to collapse in finite time (\emph{cf.} \cite{Grotto2022a}), 
that is solutions of \cref{eq:pv} for which different positions $x_i$'s coincide at positive time 
thus making the equation lose meaning as $|K(x)|\sim |x|^{-1}$ as $|x|\to 0$.
Nevertheless, the motion is almost complete with respect to natural invariant measures: 
\cref{eq:pv} are Hamilton's equations with Hamiltonian function
\begin{equation*}
    H=-\frac1{2\pi} \sum_{i\neq j} \gamma_i \gamma_j \log d(x_i,x_j)
\end{equation*}
(corresponding to the kinetic energy of the fluid, $d$ is the periodic distance on $\T$) 
in conjugate coordinates $(x_{i,1},\gamma_i x_{i,2})_{i=1}^N$ (the coordinates of single positions). 
This implies that the volume (Lebesgue) measure $dx_1\cdots dx_N$ of $\T^N$ is preserved by the dynamics, \emph{i.e.} a distribution of PVs at independent uniform random positions is an equilibrium state of the system.
It is a classical result \cite{Durr1982,Marchioro1994} that singular configurations are a negligible set of phase space 
with respect to absolutely continuous invariant measures of the system, 
that is there exists a measure-preserving flow $\Phi_t:\T^N\to\T^N$ consisting of smooth solutions  \cite{Grotto2020}.

In this paper we study the behavior of time correlations of local observables of a large number $N$ of PVs 
under the invariant measure $dx_1\cdots dx_N$.
We measure, in a numerical simulation of the system obtained through a 4th order Runge-Kutta method, local observables of the form
\begin{equation}\label{eq:observable}
    F_\sigma^L(t)=\sum_{i=1}^N \gamma_i\phi_\sigma^L(x_i(t)),
\end{equation}
where
\begin{equation*}
    \phi_\sigma^L(x)=\frac1{2\pi\sigma^2}\exp(-d(x,(L,0))^2/2\sigma^2)
\end{equation*}
is a function of $\T$ concentrated around the point $(L,0)$, the shape of which is not relevant.
In the latter, $L\geq 0$ and $d(\cdot,\cdot)$ denotes the (periodic) distance on $\T$.
We focus on the case in which $N=1000$ PVs have intensities
$\gamma_i=\pm1$, half of each sign, so that the observables under consideration are zero averaged,
$0=\brak{F_\sigma^L(0)}=\brak{F_\sigma^L(t)}$, $t>0$; 
here and in the following brackets denote integration with respect to initial data with distribution $dx_1\cdots dx_N$.
We consider local averages of the vorticity distribution $\sum_{i=1}^N \gamma_i\delta_0(x-x_i(t))$, instead of observables of the velocity field $u$.
Indeed, since the velocity field induced by a PV configuration is singular at vortices' positions, measurements on $u$ are subject to strong fluctuations and therefore less suitable to statistical analysis. On the other hand, let us stress that $u$ is recovered from the vorticity field by means of a linear operation (Biot-Savart law) performed at fixed time, and time dependence of correlations should not be influenced by such transformation.
The correlation
\begin{equation}\label{eq:rhodef}
    \rho_\sigma^L(t)=\frac{\brak{F_\sigma^L(0)F_\sigma^L(t)}}{\sqrt{\brak{F_\sigma^L(0)^2}\brak{F_\sigma^L(t)^2}}},
\end{equation}
provides for $L=0$ the autocorrelation of a single observable at times $0$ and $t$, and that of two observables for $L>0$.

Our measurement provides robust statistical evidence for a power law decay of correlations
\begin{equation}\label{eq:decay}
    \rho^L_\sigma(t)\sim \frac{1}{t},\quad t>0,
\end{equation}
the exponent being independent of the parameters of the system, \emph{i.e.} the width $\sigma$ of $\phi_\sigma^L$, 
the distance $L$ between observables ---in particular the result holds for self-correlations and correlations between distinct observables---,
the timestep of the integration scheme $\delta t$ and the regularization parameter $\epsilon$ required to perform the numerical simulation.
In particular, we considered increasingly small values of $\delta t, \epsilon$ to validate the claim for the theoretical system. 
However, let us emphasize that the range of validity of \eqref{eq:decay} in $t$ strongly depends on the choice of parameters $\sigma, L$:
larger and closer observables require a longer transient before they begin to decay. 
Moreover, the time threshold after which $\rho^L_\sigma(t)$ is too close to $0$ to be measurable in our simulation,
because of the numerical error prevailing, is smaller for smaller $\sigma>0$ and larger $L\geq 0$.

We expect our result to hold in a limiting $N\to \infty$ regime if intensities scale as $\gamma_i\sim 1/\sqrt N$, 
that is for flows of 2D Euler equations preserving the Gaussian \emph{enstrophy measure},
Gaussianity at fixed time being a product of the Central Limit scaling and equidistribution and independence of PV positions
(scaling $\gamma_i\sim 1/N$ would lead to a trivial stationary solution of Euler's equations).
A series of recent (theoretical) contributions \cite{flandoli,grotto1,grottoromito1,grottoromito2,grottopeccatiTPMS} 
has established existence of limiting measure-preserving solutions as $N\to \infty$
in the form of (analytically very weak) solutions of 2D Euler equations preserving Gaussian Energy-Enstrophy ensembles.
As observed in \cite{grottofluctuationpreprint}, the nonlinearity of the dynamics makes multi-time marginals non-Gaussian. 
In particular, even a precise control of the correlations we study in this paper would not completely characterize their distribution,
and the description of the necessarily non-Gaussian multi-marginals of equilibrium flows at different times remains an open problem.
Producing robust statistics supporting extrapolation in $N\to \infty$ is computationally intensive, 
in our experiment we have considered the evolution of systems consisting in up to $N=1000$ PVs.

Despite considerable efforts, the temporal structure of fluid mechanical models is in general only partially understood, 
and only few theoretical results are available for equilibrium flows of 2D Euler dynamics or its regularizations.
We refer to \cite{Shnirelman2013,Khesin2023} for an overview on relaxation towards simple states in 2D Euler equations,
and to \cite{Elgindi2022} concerning (spatially) scale-invariant dynamics.
Time correlations of equilibrium dynamics in closely related dynamical systems have been the object of fundamental works \cite{Alder1970,Alder1971}.
The kinetic approach to relaxation in PV dynamics has been the subject of a series of works by Chavanis \cite{Chavanis1998,Chavanis2001,Chavanis2002,Chavanis2007,Chavanis2014,Chavanis2023EPJ},
mostly focusing on the dynamics of a tracer vortex in a large ensemble:
diffusive relaxation towards equilibrium of the whole vortex system was discussed in \cite[Section 3.2]{Chavanis2012},
in which it is observed that characteristic time of relaxation does not just depend on $N$,
and it is influenced by the whole initial distribution of PVs, the one we consider not being included in the discussion;
indeed, relaxation might not even take place \cite{Chavanis2012physa}.
Describing collective effects in PV dynamics is in general a difficult task \cite{Chavanis2008}:
to the best of our knowledge, our result is the first evidence of power law decay of correlation in large PV systems.

Ergodicity of the PV system was conjectured by Onsager \cite{Onsager1949} and disproved by Khanin \cite{Khanin1982} in a system of few PVs.
Our results however show that the equilibrium state corresponding to the invariant measure $dx_1\cdots dx_N$ of \Cref{eq:pv} for a system of many PVs 
exhibits mixing behavior: notwithstanding the existence of singular solutions and (quasi-)periodic orbits \cite{Lim1989,Blackmore2005} this suggests that in the limit $N\to \infty$ they become an increasingly smaller set of phase space, and ergodicity might be recovered. 
Indeed, the latter was proposed by \cite[p. 865]{Eyink1993} as a condition much weaker than strict ergodicity that justifies the statistical equilibrium assumption for PV dynamics, while the rate of approach to equilibrium in ergodic components was indicated as an important issue in the scope of Onsager's theory.

The equilibrium state of the PV system under consideration 
---or Gibbsian ensembles absolutely continuous with respect to it---
is not suited for describing the vortex aggregation in turbulent flows as portrayed by Onsager,
but it might constitute a valid model for small, unresolved scales of more complex fluid dynamics systems.
It is worth mentioning that integrable and non-integrable behaviors of PV dynamics as Hamiltonian systems are the object of extensive literature, usually considering systems of a small number of PVs: we refer to the recent \cite{Modin2021} for a survey. 
Finally, while this contribution is concerned with aspects of PV systems oriented to the description of classical 2D fluids, 
let us recall that PV systems are also widely employed in 2D cold-atom systems, \emph{e.g.} Bose gas \cite{Maestrini2019,Lydon2022,Skipp2023}.

\section{Point Vortex Dynamics}

We consider a system of $N$ identical PVs with intensities $\pm1$, half of each sign,
evolving on $\T$ according to the dynamics of \cref{eq:pv}.

\subsection{Desingularization of the Interaction Kernel}

The advecting vector field in \cref{eq:pv} can be represented
via the Fourier series expansion
\begin{equation}
    K(x)= \frac1{4\pi}\sum_{k\in\Z^2_0} \frac{ik^\perp e^{i\pi k\cdot x}}{|k|^2},\quad x\in\T.
\end{equation}
The latter is singular at $x\to 0$, as it is the (orthogonal) gradient of the Green function $G$ of $\T$,
which has a logarithmic singularity.
In fact, the periodic boundary does not affect asymptotic behavior of $G$ at $x\to 0$,
which is the same of the free Green function $-\frac1{2\pi}\log|x|$.
Because of the relatively large number of PVs, the relevant interactions are those of close PVs \cite{Chavanis2002},
so that for computational aims we can safely replace $K$ with (a regularization of) the orthogonal gradient of the free Green function.

Numerical simulations of PV dynamics have been performed using the regularized interaction kernel
\begin{equation*}
    K_\epsilon(x,y)=\frac{(y_2-x_2,x_1-y_1)}{2\pi(d(x,y)+\epsilon)^2},
\end{equation*}
where $d(x,y)$ denotes the periodic distance on the 2D torus $\T=[-1,1]^2$.
As we just mentioned, even at the ideal value $\epsilon=0$ the kernel still deviates from the original $K$:
the error is negligible at close distances ($|x|<0.1$) and overall bounded by $|K-K_0|< 0.2$, 
the latter being negligible compared to the singularity at $x=0$.
Polynomial corrections of $K_0$ allow to improve the approximation arbitrarily. 
However, testing has revealed that already a quadratic correction produces negligible fluctuations in the quantities we have measured.
Again, this is because in the statistical ensemble under consideration the most relevant interactions of PVs take place at close distances.
On the other hand, spectral methods, \emph{i.e.} Fourier truncation of the fluid velocity (thus of $K$), are not suited for our purposes: 
they approximate well interactions of distant PVs, but become computationally inefficient in approximating the more relevant and singular close-range interactions.

\begin{figure*}
  \setlength{\unitlength}{\textwidth}
    \includegraphics[width=.48\textwidth]{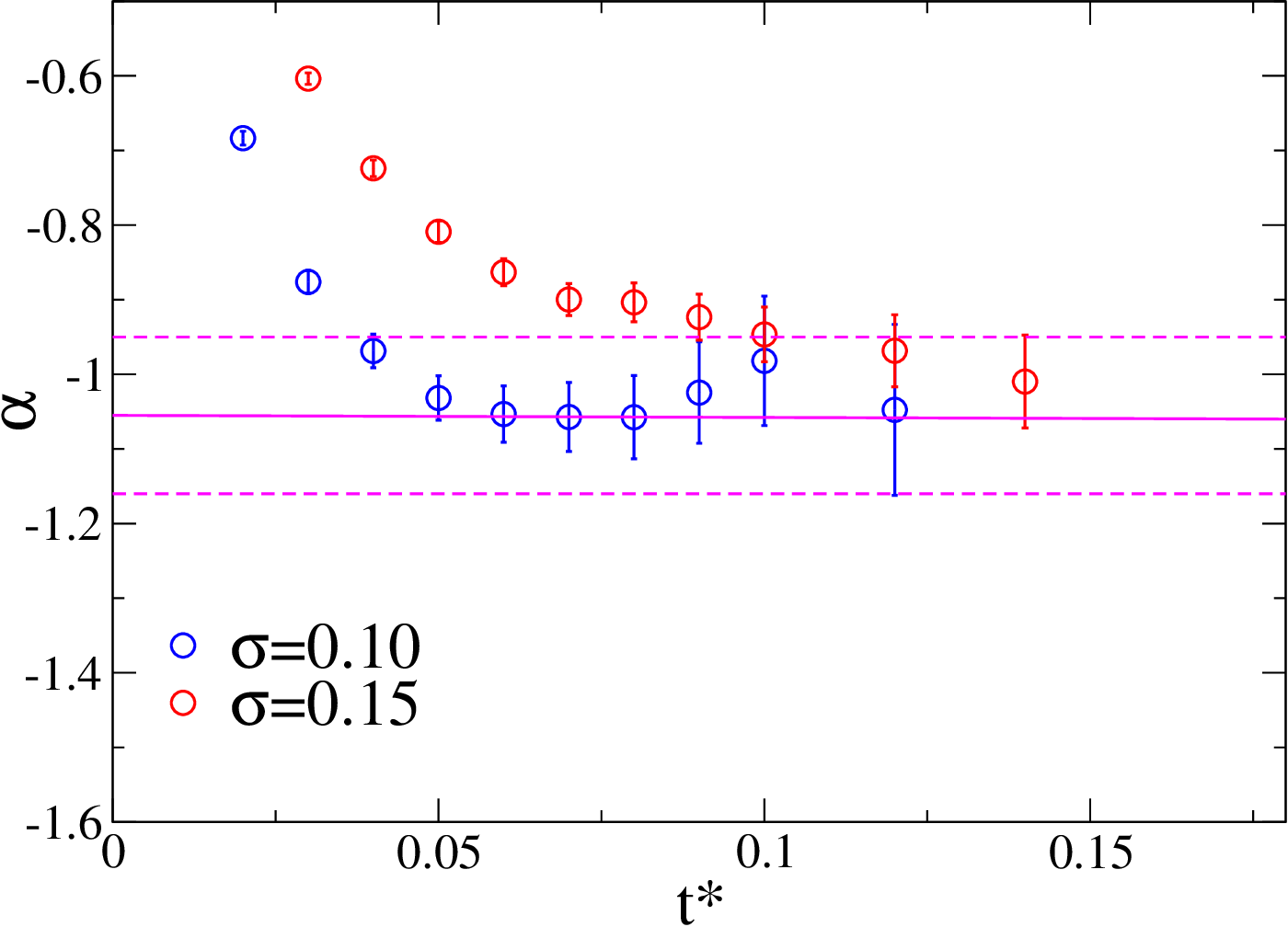}
    \includegraphics[width=.48\textwidth]{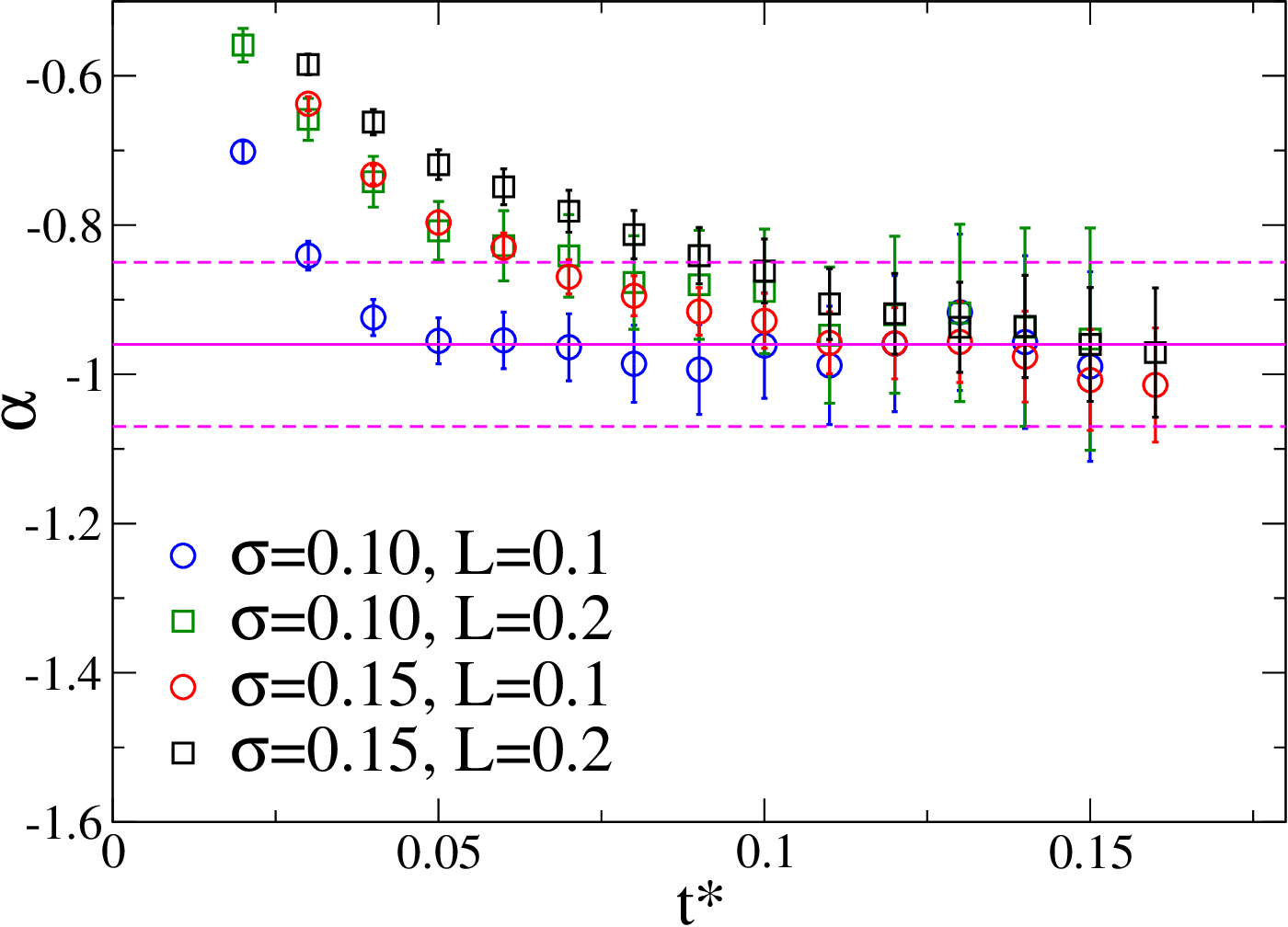}
\caption{\label{fig:fit} Estimates of the power law exponent $\alpha$, computed by fitting the logarithm of $\rho^L_\sigma(t)$ for different values of $\sigma$ and various different fitting ranges $[t^\ast, t^{\ast\ast}]$. Horizontal bands denotes the final confidence interval $\alpha=-1.06(11)$ (obtained by taking into account fit systematics) for the self-correlation coefficient, $L=0$, and $\alpha=-1.02(14)$ for the correlation coefficient, $L>0$, the two values being compatible.}
\end{figure*}

\begin{figure*}
  \setlength{\unitlength}{\textwidth}
    \includegraphics[width=.48\textwidth]{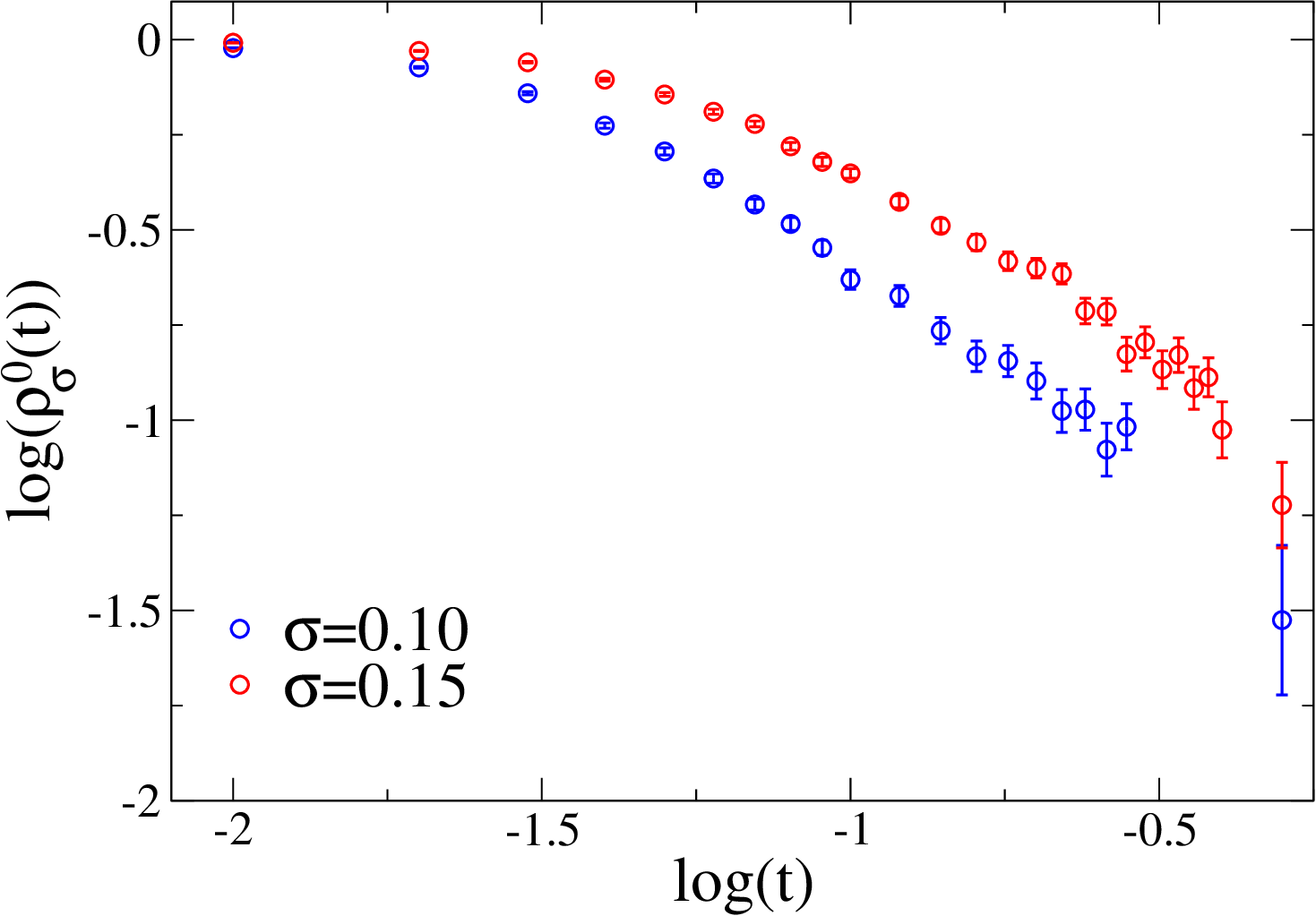}
    \includegraphics[width=.48\textwidth]{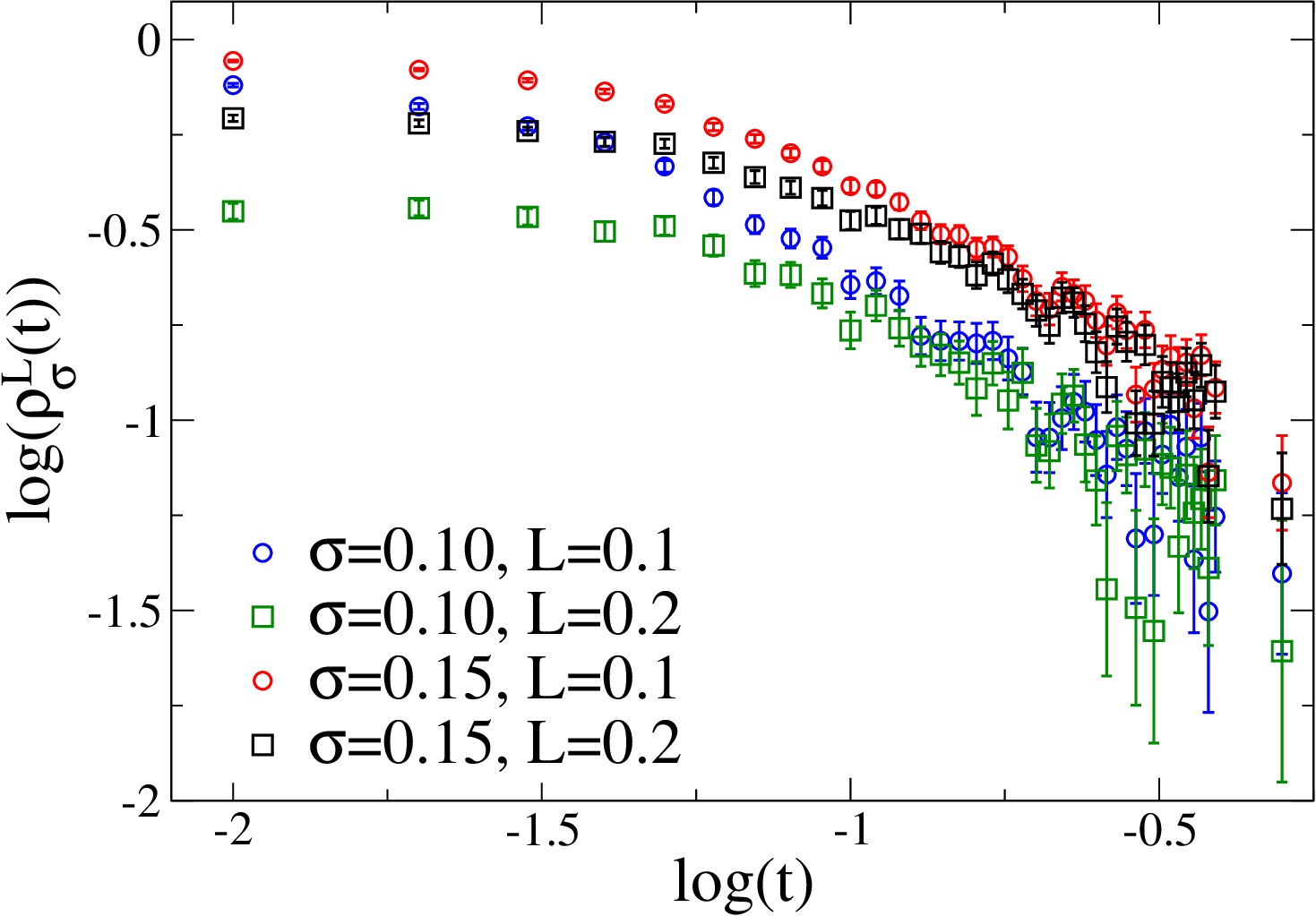}
\caption{\label{fig:plot} 
Logarithm of $\rho^L_\sigma(t)$ as a function of the logarithm of $t$; $\rho^{L=0}_\sigma(t)$ represents the Pearson autocorrelation ($L=0$, on the left) and correlation ($L>0$, on the right) coefficient between the observables $F(0,0, \phi(\sigma))$ and $F(t,L, \phi(\sigma))$. The dynamics is integrated with timestep $\delta t=10^4$ and machine $\epsilon$ regularization, for different values of $\sigma$ and $L$.}
\end{figure*}

\subsection{Accuracy of Numerical Approximations}

The dynamics in \cref{eq:pv} was numerically solved by means of 4th order Runge-Kutta method.
In order to evaluate the robustness of our measure, 
we considered different values of the timestep $\delta t= 10^{-3}, 10^{-4}, 10^{-5}$
and of the regularization parameter $\epsilon=10^{-3},10^{-6},10^{-16}$,
the latter value reaching machine epsilon.

For each choice of parameters $\delta t, \epsilon$
the system was solved up to time $T=1$, starting from an initial configuration of $N=1000$ i.i.d. uniformly distributed PV.
The relevant time frame in which the decay curve was observed always remained well below $T$, 
and in most cases already at half time correlations were close to zero so that their fluctuations could not be distinguished from numerical error.
Numerical integration was performed for at least $10^4$ samples of the initial configuration for each choice of 
$\delta t, \epsilon$, ensuring statistical robustness and independent sampling.

In order to gauge numerical accuracy we employed the energy (Hamiltonian) of the regularized PV system,
\begin{equation*}
    H_\epsilon=-\frac1{2\pi} \sum_{i\neq j} \gamma_i \gamma_j \log\pa{d(x_i,x_j)+\epsilon},
\end{equation*}
which is a first integral of the regularized motion. 
Numerical integration of PV dynamics with any of the above choices of parameters $\delta t, \epsilon$ 
led to a relative error $H_\epsilon(t)/H_\epsilon(t=0)$ of order at most $10^{-7}$ for time $0\leq t\leq 0.5$,
uniformly with respect to the (random) initial conditions under consideration.

\section{Time Decay of Correlations}

\subsection{Sampling procedure}
Most of the sampling was performed with $\delta t = 10^{-4}$ and $\epsilon = 10^{-16}$, as we checked that a smaller timestep ($\delta t = 10^{-5}$) 
or a larger regularization parameter lead to results compatible with the ones obtained with the adopted numerical setup.

Correlations $\rho^L_\sigma(t)$ were measured by the standard Pearson estimator. Exponential decay was ruled out because of large chi-square score of the fit, then the exponent $\alpha$ of the power law parametric model $C t^\alpha$ for $\rho^L_\sigma(t)$ 
was estimated via least square error fit of the logarithm $\log \rho^L_\sigma(t)$ (with $C$ remaining a free parameter).
In \cref{fig:plot} data points for $\log \rho^L_\sigma(t)$ are shown in the case of self-correlation $L=0$ and correlation between distinct observables $L>0$.
The linear fitting procedure was carried out for different choices of the parameters $\sigma$ and $L$,
results are reported in \cref{tab:table1,tab:table2,tab:table3}.
For all the simulations, data points computed at different times and parameters are independent from each other, as they have been estimated using different samples; statistical errors were estimated by means of bootstrap procedures.

\subsection{Time dependence}
For any choice of the parameters, $\rho^L_\sigma(t)$ starts at $\rho^L_\sigma(0)=1$ and drops to values close to 0 after a short transient phase.
The fitting range $[t^\ast,t^{\ast\ast}]$ has to be chosen properly: $t^\ast>0$ must be large enough so to neglect the transient phase and $t^{\ast\ast}>t^\ast$
should not be too large, since when $\rho^L_\sigma(t)$ is too close to 0 its variations can not be distinguished from numerical error.
The choice of $t^\ast$ was the more sensitive: \cref{tab:table1} reports the results for increasing values of $t^\ast$,
and shows how the inclusion on the transient for small values of $t^\ast$ makes the fit unstable, as revealed by much higher $\chi^2$-scores.
In our simulations, variations of $t^{\ast\ast}$ (in the simulated time interval) did not change 
the result of the fit except for a slightly larger $\chi^2$-score for larger $t^{\ast\ast}$.
Dependence on fitting range --fit parameter $\alpha$ against lower extremum $t^\ast$-- is illustrated in \cref{fig:fit}.

\subsection{Results and dependence on parameters}
Simulations were run with different values of the width $\sigma$ of $\phi^L_\sigma$; 
we report only results obtained for $\sigma = 0.1$ and $\sigma = 0.15$, as smaller values of $\sigma$ 
lead to a shorter time frame in which $\rho^L_\sigma(t)$ drops from values close to 1 to values close to 0, 
thus making curve fitting procedures less stable.

Concerning the dependence on the timestep $\delta t$ and the regularization parameter $\epsilon$,
the (negligible) deviations from the reference case $\delta t=10^{-4}$ and $\epsilon = 10^{-16}$
for other choices of those parameters are collected in \cref{tab:table2,tab:table3}.

Considering the systematic errors that arise from the variable fitting range, 
we ultimately estimate $\alpha = -1.06(11)$ for the case $L=0$ and $\alpha = -1.02(14)$ for the other choices of $L$, 
the two estimates for $\alpha$ being fully compatible with $-1$.

\subsection{Dependence on the number $N$ of vortices}

Correlations were measured with the same procedure in systems of $N=10$ and $N=100$ PVs,
rescaling intensities $\gamma_i\sim 1/\sqrt N$ in order to have comparable characteristic scales of decay, \cref{tab:otherN} reports the result.
While the results for $N=10$ vortices are not compatible with those with $N=1000$ (the former exhibit slower decay rate),
results for $N=100$ are compatible with the power law rate \cref{eq:decay}:
taking into account the relative error this provides moderate evidence for the robustness of $1/t$ time decay at larger $N$.
Errors for the fit at $N=100$ and $N=1000$ are of comparable magnitude: even if the case $N=1000$ is closer to the ideal $N\to\infty$
regime the numerical error due to the large number of interactions makes a precise fit harder to obtain.


\begin{table*}
\caption{Results of the fit of the angular coefficient $\alpha$ for $\log(\rho^L_\sigma(t))$ data, with $\delta t = 10^{-4}$ and $\epsilon = 10^{-16}$.
We report the estimated values of $\alpha$ and the reduced $\chi^2$-score of the fit, together with the degrees of freedom.}
\label{tab:table1}
\begin{ruledtabular}
\begin{tabular}{c|ccc|ccc}
 &\multicolumn{3}{c|}{$L=0$, $\sigma=0.1$} &\multicolumn{3}{c}{$L=0$, $\sigma=0.15$}\\
 $t^\ast$
 &$\alpha$    & $\chi^2$/dof & dof
 &$\alpha$    & $\chi^2$/dof & dof\\ 
 \hline
    0.02   &   -0.684(9)  & 17.4   &   17 
    &   ---    &    ---  &    --- \\
    0.03   &   -0.876(15)   &   3.0  &   16
    &  -0.604(8)    &    16.4  &    22 \\
    0.04   &   -0.969(23)  &    1.1  &    15 
    &   -0.724(11)   &   6.7   &   21 \\
    0.05    &  -1.032(30)  &  0.5   &   14 
    &   -0.809(14)    &    2.7  &    20 \\
    0.06    &  -1.053(38)  &  0.4 &      13 
    &  -0.863(18)   &   1.5  &   19 \\
    0.07    &  -1.057(46)  &   0.4 &      12 
    &   -0.900(21)    &    1.0  &    18 \\
    0.08    &  -1.058(56)  &    0.5 &      11
    &  -0.904(26)   &   1.1   &   17 \\
    0.09   &   -1.025(68)   &   0.5  &   10
    &  -0.923(31)   &   1.1   &   16 \\
    0.10   &   -0.982(87)   &   0.4  &   9
    &  -0.947(37)   &   1.1   &   15 \\
    0.12   &   -1.048(115)   &   0.4  &   8
    &  -0.969(48)   &   1.1   &   14 \\
    0.14   &   ---   &   ---  &   ---
    &  -1.010(62)    &  1.1    &  13 \\
 \hline
 &\multicolumn{3}{c|}{$L=0.1$, $\sigma=0.1$} &\multicolumn{3}{c}{$L=0.1$, $\sigma=0.15$}\\

 $t^\ast$
 &$\alpha$    & $\chi^2$/dof & dof
 &$\alpha$    & $\chi^2$/dof & dof\\ 
 \hline
0.02 	& -0.702(14) &	 4.6 &	 37 
&   ---   &   ---  &   --- \\
0.03 	& -0.841(19) &	 1.6 &	 36 
& -0.638(10) & 	 6.9 & 	 36 \\
0.04 	& -0.924(24) &	 0.7 &	 35 
& -0.733(12) & 	 2.8 & 	 35 \\
0.05 	& -0.955(31) &	 0.7 &	 34 
& -0.797(16) & 	 1.5 & 	 34 \\
0.06 	& -0.955(38) &	 0.7 &	 33 
& -0.830(20) & 	 1.3 & 	 33 \\
0.07 	& -0.964(45) &	 0.7 &	 32 
& -0.869(23) & 	 1.0 & 	 32 \\
0.08 	& -0.986(52) &	 0.7 &	 31 
& -0.895(27) & 	 1.0 & 	 31 \\
0.09 	& -0.994(60) &	 0.7 &	 30 
& -0.916(32) & 	 0.9 & 	 30 \\
0.10 	& -0.961(70) &	 0.7 &	 29 
& -0.928(37) & 	 0.9 & 	 29 \\
0.11 	& -0.988(79) &	 0.7 &	 28 
& -0.958(41) & 	 0.9 & 	 28 \\
0.12 	& -0.959(91) &	 0.8 &	 27 
& -0.959(48) & 	 0.9 & 	 27 \\
0.13 	& -0.917(105) &	 0.8 &	 26 
& -0.957(54) & 	 1.0 & 	 26 \\
0.14 	& -0.957(116) &	 0.8 &	 25 
& -0.976(61) & 	 1.0 & 	 25 \\
0.15 	& -0.990(127) &	 0.8 &	 24 
& -1.008(68) & 	 1.0 & 	 24 \\
0.16 &   ---   &   ---  &   --- 
& -1.014(76) & 	 1.0 & 	 23 \\
\hline
&\multicolumn{3}{c|}{$L=0.2$, $\sigma=0.1$}&\multicolumn{3}{c}{$L=0.2$, $\sigma=0.15$}\\
 $t^\ast$
 &$\alpha$    & $\chi^2$/dof & dof
 &$\alpha$    & $\chi^2$/dof & dof\\ 
 \hline 
0.02 	& -0.559(23) &	 2.4 &	 37 
&   ---   &   ---  &   --- \\
0.03 	& -0.658(28) &	 1.5 &	 36 
& -0.585(14) &	 3.9 & 	 36 \\
0.04 	& -0.742(34) &	 1.0 &	 35 
& -0.662(17) &	 2.3 & 	 35 \\
0.05 	& -0.808(39) &	 0.7 &	 34 
& -0.719(20) &	 1.5 & 	 34 \\
0.06 	& -0.828(47) &	 0.7 &	 33 
& -0.749(24) &	 1.4 & 	 33 \\
0.07 	& -0.841(55) &	 0.7 &	 32 
& -0.781(28) &	 1.3 & 	 32 \\
0.08 	& -0.877(63) &	 0.7 &	 31 
& -0.813(32) &	 1.2 & 	 31 \\
0.09 	& -0.880(73) &	 0.7 &	 30 
& -0.841(38) &	 1.1 & 	 30 \\
0.10 	& -0.889(84) &	 0.7 &	 29 
& -0.862(43) &	 1.1 & 	 29 \\
0.11 	& -0.948(91) &	 0.6 &	 28 
& -0.906(48) &	 1.0 & 	 28 \\
0.12 	& -0.920(105) &	 0.7 &	 27 
& -0.919(54) &	 1.0 & 	 27 \\
0.13 	& -0.918(119) &	 0.7 &	 26 
& -0.937(60) &	 1.1 & 	 26 \\
0.14 	& -0.937(133) &	 0.7 &	 25 
& -0.936(69) &	 1.1 & 	 25 \\
0.15 	& -0.953(149) &	 0.7 &	 24 
& -0.960(76) &	 1.1 & 	 24 \\
0.16 &   ---   &   ---  &   --- 
& -0.971(87) &	 1.2 & 	 23 \\
\end{tabular}
\end{ruledtabular}
\end{table*}


\begin{table*}
\caption{Correlation $\rho^L_\sigma(t)$ computed for different values of $\delta t$; the difference between the results obtained in the setup used for the majority of the simulations ($\delta t = 10^{-4}$) and the results obtained for different values of $\delta t$ is reported, for different choices of the parameter $\sigma$ and different values of $L$. The regularization parameter is taken as $\epsilon = 10^{-16}$.}
\label{tab:table2}
\begin{ruledtabular}
\begin{tabular}{c|c|c}
 &\multicolumn{1}{c|}{$L=0$, $\sigma=0.1$}&\multicolumn{1}{c}{$L=0$, $\sigma=0.15$}\\
 $\delta t$
 &$\rho^L_\sigma(t=0.08, \delta t)-\rho^L_\sigma(t=0.08,\delta t=10^{-4})$ 
 &$\rho^L_\sigma(t=0.08, \delta t)-\rho^L_\sigma(t=0.08,\delta t=10^{-4})$ \\ 
 \hline
    $10^{-5}$   &   0.004(16) 
    &   -0.012(15)     \\
    $10^{-3}$   &  0.037(16) 
    &  -0.005(14)  \\
 \hline
 &\multicolumn{1}{c|}{$L=0.1$, $\sigma=0.1$}&\multicolumn{1}{c}{$L=0.1$, $\sigma=0.15$}\\
 $\delta t$
 &$\rho^L_\sigma(t=0.08, \delta t)-\rho^L_\sigma(t=0.08, \delta t=10^{-4})$ 
 &$\rho^L_\sigma(t=0.08, \delta t)-\rho^L_\sigma(t=0.08, \delta t=10^{-4})$\\ 
 \hline
    $10^{-5}$   &   -0.000(20)
    &   -0.008(17)     \\
    $10^{-3}$   &  0.0436(20)
    &  0.002(17)    \\
\hline
&\multicolumn{1}{c|}{$L=0.2$, $\sigma=0.1$}&\multicolumn{1}{c}{$L=0.2$, $\sigma=0.15$}\\
 $\delta t$
 &$\rho^L_\sigma(t=0.08, \delta t)-\rho^L_\sigma(t=0.08, \delta t=10^{-4})$ 
 &$\rho^L_\sigma(t=0.08, \delta t)-\rho^L_\sigma(t=0.08, \delta t=10^{-4})$\\  
 \hline 
    $10^{-5}$   &   -0.004(21)
    &   -0.018(19)     \\
    $10^{-3}$   &  0.009(21)
    &  -0.020(18)   \\
\end{tabular}
\end{ruledtabular}
\end{table*}


\begin{table*}
\caption{Correlation $\rho^L_\sigma(t)$ computed for different values of $\epsilon$; the difference between the results obtained in the setup used for the majority of the simulations ($\epsilon = 10^{-16}$) and the results obtained for different values of $\epsilon$ is reported, for different choices of the parameter $\sigma$ and different values of $L$. The timestep is taken as $\delta t = 10^{-4}$.}
\label{tab:table3}
\begin{ruledtabular}
\begin{tabular}{c|c|c}
 &\multicolumn{1}{c|}{$L=0$, $\sigma=0.1$}&\multicolumn{1}{c}{$L=0$, $\sigma=0.15$}\\
 $\epsilon$
 &$\rho^L_\sigma(t=0.08,\epsilon)-\rho^L_\sigma(t=0.08,\epsilon=10^{-16})$ 
 &$\rho^L_\sigma(t=0.08,\epsilon)-\rho^L_\sigma(t=0.08,\epsilon=10^{-16})$\\ 
 \hline
    $10^{-6}$   &   0.001(16) 
    &    -0.015(15)      \\
    $10^{-3}$   & -0.002(16)
    &  -0.012(15)  \\
 \hline
 &\multicolumn{1}{c|}{$L=0.1$, $\sigma=0.1$}&\multicolumn{1}{c}{$L=0.1$, $\sigma=0.15$}\\
 $\epsilon$
 &$\rho^L_\sigma(t=0.08,\epsilon)-\rho^L_\sigma(t=0.08,\epsilon=10^{-16})$ 
 &$\rho^L_\sigma(t=0.08,\epsilon)-\rho^L_\sigma(t=0.08,\epsilon=10^{-16})$\\ 
 \hline
    $10^{-6}$   &   0.018(20)
    &   0.006(17)     \\
    $10^{-3}$   &  0.003(20)
    &  0.002(17)    \\
\hline
&\multicolumn{1}{c|}{$L=0.2$, $\sigma=0.1$}&\multicolumn{1}{c}{$L=0.2$, $\sigma=0.15$}\\
 $\epsilon$
 &$\rho^L_\sigma(t=0.08,\epsilon)-\rho^L_\sigma(t=0.08,\epsilon=10^{-16})$ 
 &$\rho^L_\sigma(t=0.08,\epsilon)-\rho^L_\sigma(t=0.08,\epsilon=10^{-16})$\\ 
 \hline 
    $10^{-6}$   &   0.051(21)
    &   0.022(20)     \\
    $10^{-3}$   &  0.011(21) 
    & -0.003(19)   \\
\end{tabular}
\end{ruledtabular}
\end{table*}




\begin{table*}\label{tab:otherN}
\caption{Results of the fit of the angular coefficient $\alpha$ for $\log(\rho^{L=0}_{\sigma}(t))$ data, with $\delta t = 10^{-4}$ and $\epsilon = 10^{-16}$.
We report the estimated values of $\alpha$ and the reduced $\chi^2$-score of the fit, together with the degrees of freedom, for different values of $N$ and $\sigma$.}
\label{tab:tableN_bis}
\begin{ruledtabular}
\begin{tabular}{c|ccc|ccc}
 &\multicolumn{3}{c|}{$N=10$, $L=0$, $\sigma=0.1$} &\multicolumn{3}{c}{$N=10$, $L=0$, $\sigma=0.15$}\\
 $t^\ast$
 &$\alpha$    & $\chi^2$/dof & dof
 &$\alpha$    & $\chi^2$/dof & dof\\ 
 \hline
0.02 &	 -0.674(12)	 &	 8.7 &	 17 
&   ---   &   ---  &   --- \\
0.03 &	 -0.820(18) &	 2.2 &	 16 
&  -0.562(9) &	 12.8 &	 22 \\
0.04 &	 -0.908(25) &	 0.6 &	 15 
& -0.670(12) &	 4.9 &	 21 \\
0.05 &	 -0.932(32) &	 0.5 &	 14 
& -0.743(15) &	 2.2 &	 20 \\
0.06 &	 -0.923(40) & 	 0.5 &	 13 
& -0.794(18) &	 0.9 &	 19 \\
0.07 &	 -0.920(48) &	 0.6 &	 12 
& -0.811(22)  &	 0.8 &	 18 \\
0.08 &	 -0.899(57) &	 0.6 &	 11 
& -0.825(26) &	 0.8 &	 17 \\
0.09 &	 -0.920(65) &	 0.6 &	 10 
& -0.850(30) &	 0.7 &	 16 \\
0.10 &	 -0.849(82) &	 0.5 &	 9
& -0.843(36) &	 0.7 &	 15 \\
0.12 &	 -0.877(109) &	 0.5 &	 8 
& -0.860(46) &	 0.8 &	 14 \\
0.14 &   ---   &   ---  &   --- 
&   -0.898(59) &	 0.7 & 13 \\
\hline
 &\multicolumn{3}{c|}{$N=100$, $L=0$, $\sigma=0.1$} &\multicolumn{3}{c}{$N=100$, $L=0$, $\sigma=0.15$}\\
 $t^\ast$
 &$\alpha$    & $\chi^2$/dof & dof
 &$\alpha$    & $\chi^2$/dof & dof\\ 
 \hline
0.02 &	 -0.678(11) &	 11.3 &	 17 
&   ---   &   ---  &   --- \\
0.03 &	-0.859(19) &	 3.5 &	 16 
&  -0.587(9) &	 12.5 &	 22 \\
0.04 &	 -0.973(27) &	 1.3 &	 15 
&  -0.710(13) &	 5.0 &	 21 \\
0.05 &	-1.036(35) &	 0.9 &	 14
& -0.795(17) &	 2.0 &	 20 \\
0.06 &	-1.044(46) &	 1.0 &	 13
&  -0.833(21) &	 1.6 &	 19 \\
0.07 &	 -1.082(55) &	 0.9 &	 12 
&  -0.871(25) &	 1.3 &	 18 \\
0.08 &	 -1.075(67) &	 1.0 &	 11 
& -0.903(30) &	 1.1 &	 17 \\
0.09 &	 -1.124(77) &	 0.9 &	 10 
& -0.945(35) &	 0.9 &	 16 \\
0.10 &	 -0.995(99) &	 0.6 &	 9
& -0.965(43) &	 0.9 &	 15 \\
0.12 &	 -0.877(109) &	 0.5 &	 8 
& -1.053(54) &	 0.4 &	 14 \\
0.14 &   ---   &   ---  &   ---
& -1.050(70) &	 0.4 &	 13 \\
\end{tabular}
\end{ruledtabular}
\end{table*}

\section{Conclusions}

Accurate numerical simulations of PV flows allowed us to produce strong statistical evidence of a power law decay of time correlations at equilibrium, coherently with previous theoretical and numerical results on closely related models.   

The natural continuation of our study concerns two limitations of the present analysis. The first is the fixed number of PVs: the large $N$ limit should allow to extrapolate results on solutions of 2D Euler equations, but it faces the issue of rapidly increasing computational cost, which cannot be offset by reducing sample sizes without losing statistical robustness, because of the intrinsic instability of the flow. The second is the equilibrium state under consideration, which is not suited for describing turbulent phenomena. On the basis of Onsager's statistical mechanics theory, one should focus on high-energy microcanonical ensembles or negative-temperature canonical ensembles. Consistent sampling from those states presents a challenge on its own \cite{Esler2017}, so we must leave it to future studies. It is nevertheless worth observing that the system we consider might be well suited for describing small scale dynamics in turbulent flows, so that our result can be regarded as a first step in the understanding of more complex systems.


The persistence of time correlations we observed indicates that even under a relatively mixing state --that of completely independent vortex positions-- the PV system retains a certain stiffness, although aggregation phenomena cannot be observed in such an equilibrium flow, confirming the indication of \cite{Lundgren1977,Eyink1993} that scattered vortices may fail to relax towards equilibrium rapidly.
As a side note, this suggests that particular care should be taken in sampling procedures when studying numerical simulations of similar fluid dynamical models, since repeated sampling at small time intervals from the same evolution is likely to produce correlated data, unsuitable for statistical analysis.

As a final note, let us stress the fact that the model we considered describes an inviscid fluid, and viscosity can be included in the dynamics through stochastic forcing acting on single vortices \cite{Marchioro1982,grottophysicad}. The dependence of time correlations on the viscosity parameter (Reynolds number) should be closely related to anomalous dissipation effects observed in the inviscid limit of Navier-Stokes equations --which is most relevant in the study of boundary effects and fluid dynamical instabilities--
and it constitutes a further possible future extension of our study.

\begin{acknowledgments}
The authors wish to thank Giuseppe Cannizzaro and Franco Flandoli for insightful discussions and literature references. The authors are indebted to Claudio Bonati for many essential remarks concerning numerical and statistical aspects of this work.
F.G. was supported by the project \emph{Mathematical methods for climate science} funded by PON R\&I 2014-2020 (FSE REACT-EU I53C22001380006).
Numerical simulations have been performed on the Center for High Performance Computing (CHPC) at SNS.

\end{acknowledgments}

%

\end{document}